\documentclass[% 
 aip,
 pop,
 amsmath,amssymb,
reprint,%
]{revtex4-1}

\makeatletter
\newlength{\figcolwidth}
\if@twocolumn
\else
\fi
\makeatother

\usepackage{graphicx}% Include figure files
\usepackage{dcolumn}% Align table columns on decimal point
\usepackage{bm}% bold math
\usepackage[utf8]{inputenc}
\usepackage[T1]{fontenc}
\usepackage{mathptmx}
\usepackage{etoolbox}
\usepackage{siunitx}

\graphicspath{{figs/}}

\DeclareSIUnit\gauss{G}

\newcommand{\llnl}{\affiliation{Lawrence Livermore National Laboratory, Livermore CA}}
\newcommand{\MIT}{\affiliation{Massachusetts Institute of Technology, Cambridge MA}}

\makeatletter
\def\@email#1#2{%
 \endgroup
 \patchcmd{\titleblock@produce}
  {\frontmatter@RRAPformat}
  {\frontmatter@RRAPformat{\produce@RRAP{*#1\href{mailto:#2}{#2}}}\frontmatter@RRAPformat}
  {}{}
}%
\makeatother

\begin{document}

\preprint{AIP/123-QED}

%gs earliest suggests "Sustained Collisionless Interpenetrating Flows on OMEGA for Study of Nonlinear Kinetic Instability"
\title{Sustained interpenetrating plasma flows for the investigation of late time kinetic instability evolution}
% Force line breaks with \\
\author{G.D. Sutcliffe}
\email{sutcliffe2@llnl.gov}
\llnl{}
\author{N. Vanderloo}
\MIT{}
\author{C. Bruulsema}
\llnl{}
\author{V. Valenzuela-Villaseca}
\llnl{}
\MIT{}
\author{G. Swadling}
\llnl{}
\author{M. Zhou}%
\affiliation{Department of Physics and Astronomy, Dartmouth College, Hanover NH}
\author{A. Bret}%
\affiliation{ETSI Industriales, Universidad de Castilla-La Mancha, Ciudad Real, Spain}
\affiliation{Instituto de Investigaciones Energéticas y Aplicaciones Industriales, Campus Universitario de
Ciudad Real, Ciudad Real, Spain}
\author{C.K. Li}%
\MIT{}
\author{J.S. Ross}
\author{J. Moody}
\llnl{}

\date{\today}% It is always \today, today,

\begin{abstract}
Sustained collisionless interpenetrating plasma flows have been generated on the OMEGA laser facility to enable direct investigation of nonlinear evolution of fields generated by electromagnetic kinetic instabilities. FLASH simulations and Thomson scattering measurements are used to determine the plasma conditions achieved. 
Interpenetrating flows are observed to remain collisionless for at least 11 ns, longer than any prior OMEGA experiment, 
supporting the growth and nonlinear saturation of the Weibel instability. 
Resulting magnetic fields are measured using proton radiography. 
This work establishes a unique platform for late-time filament evolution measurements.
\end{abstract}

\maketitle

\section{\label{sec:level1}Introduction}

Magnetic field generation in plasmas has been a key focus of plasma physics research for many decades. An understanding of the relative importance of generation and saturation mechanisms in various regimes underpins our best understanding of magnetic fields in plasmas of diverse scales, from the microscopic sub-laser spot size scales of inertial fusion laser facilities up to galactic-scale background magnetic fields.   
% Astrophysical (interstellar, galactic background) fields are observed to be typically $\sim$\qty{10}{\micro\gauss}. 
In astrophysical plasmas, intergalactic and intra-galaxy-cluster plasmas are observed to be magnetized with $\sim$\qty{10}{\micro\gauss} fields.
One leading theory that explains the large-scale magnetism of the universe invokes initially weak seeds  produced by the Biermann battery effect, which magnetizes plasmas when density and temperature gradients are misaligned. This mechanism is estimated to produce \qty{e-20}{\gauss}-level fields, and therefore it is believed that dynamo action amplifies these seeds, reaching saturation at the observed magnetic strength \cite{kulsrud_origin_2008}.

% One such 
An alternative mechanism is the Weibel instability, which arises 
%in astrophysical and inertial fusion plasmas alike 
when the plasma distribution is anisotropic\cite{Weibel_spontaneously_1959}. 
% when there are interpenetrating flows whose interflow collisionality is too low to thermalize the distribution function \cite{Weibel_spontaneously_1959}.
This anisotropy leads to the inherently kinetic Weibel instability
% , and 
where
magnetic fields grow from
either
% other 
seeds or thermal noise. 
One common scenario involves two counter-streaming flows that can interpenetrate when the plasmas are sufficiently collisionless\cite{fried_mechanism_1959}, often referred to as the ion Weibel instability.
While Biermann fields saturate at small amplitudes, proportional to the small gradients of temperature and density, 
there are promising indications in particle-in-cell (PIC) simulations that saturated Weibel fields are insensitive to plasma tempertature/density gradient length scales \cite{schoeffler_magnetic-field_2014,schoeffler_generation_2016} and thus are candidates for seed fields that far exceed those provided by the Biermann battery.

An issue arises for the Weibel instability as a candidate seed for astrophysical fields: Weibel fields are generated at ion inertial spatial scales \cite{davidson_nonlinear_1972} which are significantly smaller than the spatial scales of intracluster/intergalactic magnetic fields.
These magnetic fields have the potential to be amplified by dynamo action, but one of the main outstanding challenges is understanding 
whether and how these microscopic fields can grow to sufficiently large coherence lengths to be picked up and amplified by large-scale turbulent flows.
In other words, how can microscopic Weibel fields contribute to macroscopic cosmic magnetic fields? 

The Weibel instability fills the volume of the interacting plasmas with ion-inertial-scale filaments, which are believed to merge in the nonlinear stages of the instability. %vicente 
The question of how small-scale fields at ion inertial scales coalesce to larger scales has been treated theoretically and with simulations \cite{medvedev_long-time_2005,zhou_magnetic_2019,ruyer_nonlinear_2015,takabe_theory_2023}. 
Theoretical models contain a range of physics, variously including\cite{zhou_magnetic_2019} or ignoring\cite{ruyer_nonlinear_2015} magnetic reconnection, which must happen to some degree as filaments merge, as a time-limiting mechanism. Other predictions include the role of the drift-kink instability to disrupt the merger evolution \cite{ruyer_disruption_2018,fiuza_electron_2020,marret_energy_2026-1}.
Comparison of experimental data to models has been limited to relatively early-time data, where the linear ion Weibel instability has saturated, but sufficient long-term evolution (which occurs on Alfvénic timescales) that would differentiate between models 
is outstanding. 
Recent efforts to diagnose these plasmas are centered around observing ion Weibel filaments and subsequent collisionless shock formation on OMEGA \cite{ross_characterizing_2012,huntington_observation_2015,park_collisionless_2015,huntington_magnetic_2017,bruulsema_local_2020,manuel_experimental_2022}, 
but not on the nonlinear evolution of filaments. 
Some efforts have occurred on other similar systems \cite{yuan_laboratory_2018, yuan_electron_2024} and NIF \cite{fiuza_electron_2020}, but data remain limited to the early phases of filament coalescence.

In this paper, we present an experimental platform that allows for the direct observation of the longest duration of ion Weibel filament mergers to date. 
We demonstrate in preliminary experiments that collisionless, low-resistivity plasma conditions are sustained for sufficiently long durations to 
distinguish between the various theoretical predictions for long-term evolution of Weibel filaments.
The paper is organized as follows: Section~\ref{sec:setup} covers the experimental setup details. Section~\ref{sec:Thomson} explains the Thomson scattering analysis. Section~\ref{sec:flash} explains the complimentary radiation-hydrodynamics simulations. Section~\ref{sec:prad} details the proton radiography measurements and derived quantities. Finally, Section~\ref{sec:discussion} discusses the implications of the measurements, and Section~\ref{sec:conclusion} concludes and outlines future work.

\section{Experimental setup}
\label{sec:setup}

\begin{figure}
\includegraphics[width=\figcolwidth]{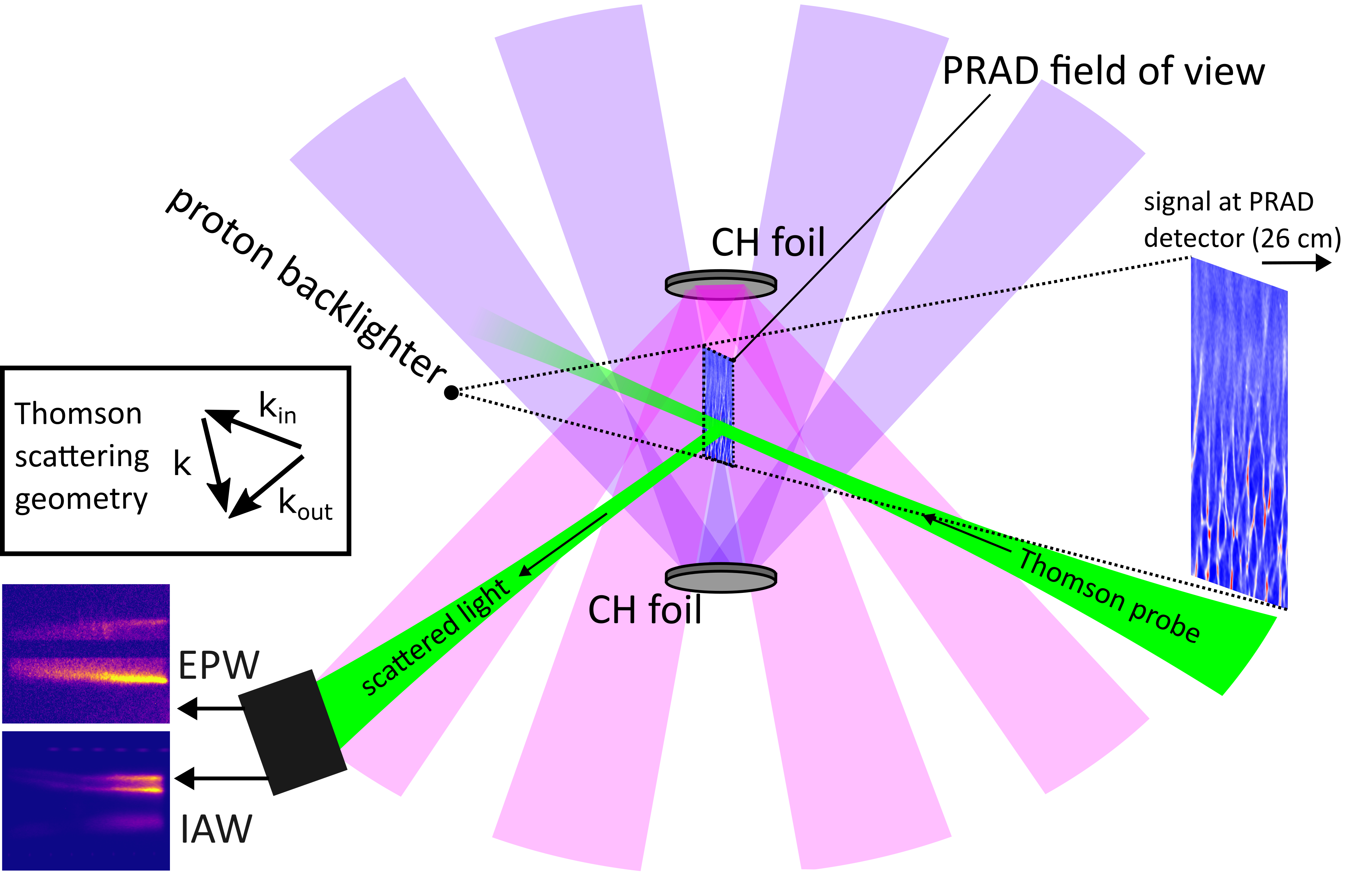}
\caption{\label{fig:expsetup} Experiment setup schematic. The upper and lower CH foils are driven by opposing beams (purple and fuchsia). The proton radiography field of view (dotted square) and the position of the backlighter are shown. The Thomson scattering probe and collected scattered light are shown in green.}
\end{figure}

Experiments were conducted at the OMEGA laser facility\cite{boehly_upgrade_1995} with the goal of sustaining and diagnosing collisionless, low-resistivity conditions in the interpenetrating plasma flows for as long as possible.
A diagram of the experiment can be seen in Fig.~\ref{fig:expsetup}. 
The experiment consists of two \qty{3}{\milli\meter} diameter circular foils with 8 mm separation.
The foils are \qty{200}{\micro\meter} polyethylene - a CH plastic with a C:H composition ratio of 1:2.
The 
front side of each foil is driven by 12 beams, 
which are temporally tiled to make pulses of \qty{6}{\nano\second} total duration; each beam has \qty{450}{J} of energy for a total per-target energy of \qty{5.4}{\kilo\joule}, power of \qty{5.4}{\tera\watt}, and intensity of \qty{2.4e14}{\watt\per\centi\meter\squared}. 
The beam circular super-Gaussian spatial profiles were set by distributed phase plates with \qty{650}{\micro\meter} diameter (1/e intensity diameter) and a super-Gaussian exponent of 2.4.
The beams employ smoothing by spectral dispersion. 

A single frequency-doubled (\qty{526.5}{nm}) beam is used as a Thomson scattering diagnostic probe. The probe is focused to a \qty{100}{\micro\meter} diameter spot at the experiment midplane. 
Scattered light at \ang{63} scattering angle is collected by telescope optics with an effective field-of-view of \qty{50}{\micro\meter}, so that the collection volume is approximately a \qty{50}{\micro\meter} diameter, \qty{100}{\micro\meter} long cylinder. 
The diagnostic is sensitive to plasma bulk flow and oscillations along the scattering vector $\mathbf{k} \equiv \mathbf{k}_{\mathrm{out}} - \mathbf{k}_{\mathrm{in}}$, where  $\mathbf{k}_{\mathrm{out}}$ and  $\mathbf{k}_{\mathrm{in}}$ are the collection and incident wave vectors respectively --- in this experiment, $k$ is \ang{31.7} relative to the flow axis of the experiment.
The scattered light is split into two channels feeding spectrometers optimized for the electron plasma wave and ion acoustic wave features, respectively. 
Measurements can be made in a time-resolved or spatially resolved configuration.
The diagnostic is run in a time-resolved configuration, where the pulse is \qty{3.7}{ns} long and has $\sim$\qty{20}{J} energy. After passing through the spectrometers, the time-resolved signals are recorded using streak cameras.

A second configuration is used for proton radiography, in which
%the leg 2 driver feeds
19 beams, each with \qty{500}{J}, \qty{1}{ns} pulses drive a D$^3$He thin-glass capsule proton backlighter \cite{li_monoenergetic_2006}. The capsule implodes and $\sim$\num{e8}-\num{e9} DD (3.0 MeV) and D$^3$He (14.7 Mev) protons are isotropically emitted from a source volume of diameter \qty{50}{\micro\meter}, setting the imaging resolution\cite{li_monoenergetic_2006}. Those that pass through the subject experiment plasma are recorded on CR-39 detectors \cite{seguin_spectrometry_2003}. Small deflections from Lorentz forces inside the plasma alter the trajectories of the protons, and signatures of these forces are encoded on the detector as variations in proton flux. Various algorithms exist to reconstruct the path-integrated deflections accumulated by the protons, which can be related to magnetic and electric fields \cite{sulman_efficient_2011-1,kasim_quantitative_2017,bott_proton_2017}.

\section{Time-resolved Thomson scattering measurements}
\label{sec:Thomson}

\begin{figure*}
    \includegraphics[width=\textwidth]{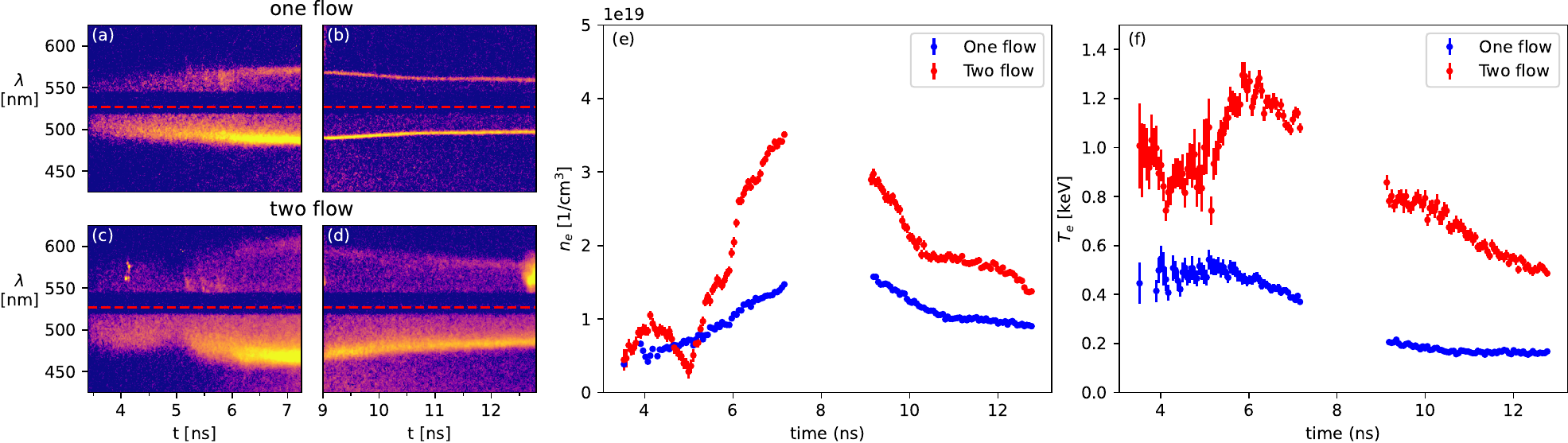}
    \caption{\label{fig:epwdata} Plasma conditions as recorded by time-resolved electron plasma wave feature of Thomson scattering. Time resolved streak data (a-d) are shown with logarithmic colorscale to highlight details at different signal levels. Fits (e-f) of $n_e(t)$ and $T_e(t)$ assuming a single Maxwellian population of electrons, for single-flow (blue) and two-flow (red) cases. Data is from OMEGA shots 112163, 112167 (one-flow) and 112165, 112168 (two-flow).}
\end{figure*}

Time-resolved Thomson scattering provides a direct, local measurement of plasma conditions in the interaction region. We use the electron plasma wave (EPW) feature to infer electron density and temperature, and the ion acoustic wave (IAW) feature to infer the ion flow velocities, ion temperatures, and ratio of flow densities.

Thomson scattering spectra in the single-flow case and two-flow cases were recorded in a time-resolved mode described above. 
The time-resolved electron plasma wave (EPW) spectra, and fitted plasma parameters, are shown in Fig.~\ref{fig:epwdata}.
The electron plasma wave in both single-flow and double-flow cases is fit to a Thomson scattering spectrum model which treats the electrons as a single maxwellian population, which is justified because the electron-ion mean free path, $\lambda^{e^{-}\backslash i}_{mfp}$, is very small, $\sim$\qty{100}{\micro\meter}. The electron population mean velocity is assumed to be zero in the lab frame which is a reasonable approximation because the flow velocities are significantly smaller than the EPW phase velocities, even for the lowest $n_e$ obsereved: for $n_e\sim$\qty{5e18}{\per\centi\meter\cubed}, $v_{\mathrm{ph,EPW}}\sim$\qty{10000}{\kilo\meter\per\second}$\gg v_{flow}\sim$\qty{1000}{\kilo\meter\per\second}. 
Background subtraction is accomplished by allowing a background function as a component of the fit model, linear in wavelength.

The fitted electron density time history is shown in Fig.~\ref{fig:epwdata}e. 
The time history is similar for both single-flow and two-flow experiments, but the two-flow electron density is approximately double single-flow case, consistent with two overlapping, noninteracting flows.  
The electron temperature history (Fig.~\ref{fig:epwdata}f) shows that some of the energy of the interpenetrating flows is being coupled to the electrons, as the two-flow electrons reach a higher temperature than their single-flow counterparts.  
Details of this coupling will be investigated in a future paper; previous studies explain some of the early-time heating of electrons through frictional heating with the ions \cite{ross_transition_2017}, while other simulation-based work indicates MHD instabilities as the source of electron heating \cite{ruyer_disruption_2018,marret_energy_2026-1}.

\begin{figure*}
\includegraphics[width=\textwidth]{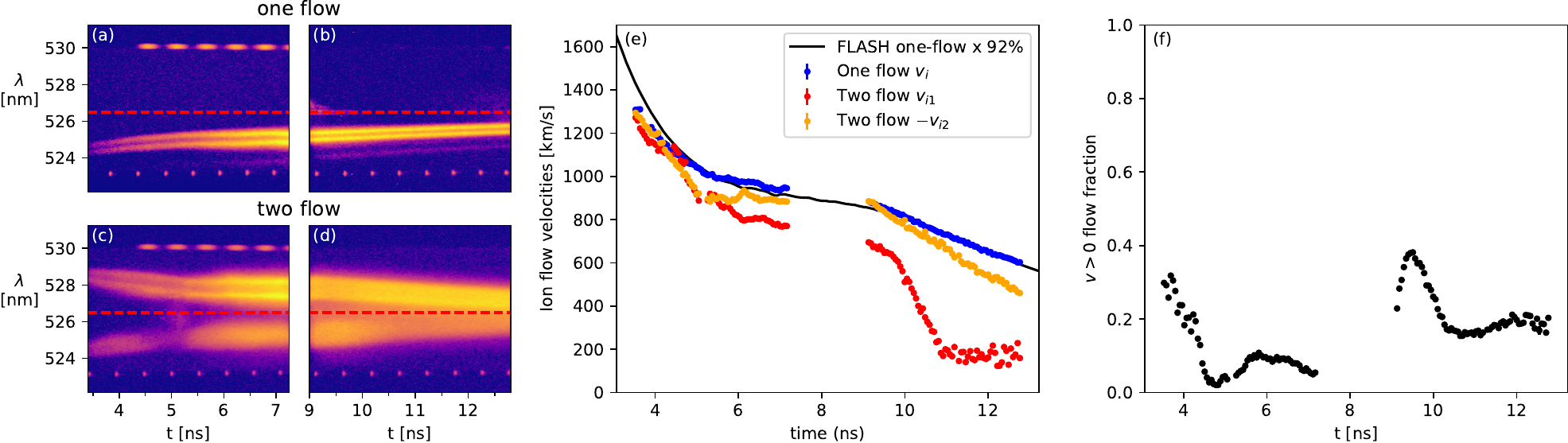}
\caption{\label{fig:iawdata} Flow conditions as recorded by time-resolved ion acoustic wave feature of the Thomson scattering data. Time resolved streak data (a-d) are shown with logarithmic colorscale to highlight low-signal details. Fits (e) of $v_i(t)$ assuming a Maxwellian populations of carbon and hydrogen ions, for single-flow (blue) and two-flow (red/orange) cases. Panel (f) shows the ion densitiy fraction of the positive-moving flow in the two-flow case. Data is from OMEGA shots 112163, 112167 (one-flow) and 112165, 112168 (two-flow).} 
\end{figure*}

The time-resolved ion acoustic wave (IAW) spectra, and fitted plasma parameters, are shown in Fig.~\ref{fig:iawdata}. This part of the Thomson scattering spectrum is primarily sensitive to details of the ion velocity distribution functions: the position of features, to first order, corresponds to flow velocity. The two-flow IAW spectra show two separate components at approximately equal but opposed velocities. That these components are spatially co-located is evidence of interpenetrating collisionless flows. 

The one-flow IAW data are fit to a model with a single flow comprised of 1:2 C:H, matching the target composition. The temperature of the C and H ions are assumed to be the same as they do not have sufficient thermal velocity to be collisionless with respect to the ions in their own flow. A similar argument applies for relative bulk velocity, so the ions are assumed to move at the same flow velocity. 

The two-flow IAW data requires a more complex model for a good fit. 
Like the EPW, the electron population is assumed to be Maxwellian with temperature $T_e$ and density $n_e$. There are now two separate ion flows with velocities $v_{i1}>0$ and $v_{i2}<0$. The fraction of the ion number density of flow 1 with respect to the total is also a fit parameter, necessary to explain the time-varying amplitudes of the IAW features. Each flow is comprised of fully ionized C and H ions, with a fixed C:H ratio. 
The absolute ion densities are determined such that the plasma is quasineutral. 
The fitted velocities of the IAW features are shown in Fig.~\ref{fig:iawdata}e, along with that from a single-flow FLASH simulation. All velocities, in both the one- and two-flow cases, are corrected for the \ang{31.7} angle between the flow and the Thomson k-vector.

% soft TODO: consider TALK ABOUT MODEL CHOICE
% soft TODO: consider SHOW MODEL FIT TO EXAMPLE LINEOUT

\section{FLASH simulations}
\label{sec:flash}

FLASH radiation-hydrodynamics simulations \cite{fryxell_flash_2000,tzeferacos_flash_2015}
were conducted as part of the experimental design process. 
The purpose of the FLASH simulations here is not to model kinetic filamentation, but to provide a hydrodynamically consistent description of the single-flow ablation that sets the upstream conditions for the counter-streaming problem. Because the collision of two high-velocity flows in this regime is inherently kinetic, the two-flow interaction itself is not captured in radiation-hydrodynamics. We therefore use FLASH only to benchmark the single-flow velocity and density evolution against Thomson scattering. 

The simulations utilized the well-adopted LaserSlab template, in which beams are incident on a thin foil of material (in this case CH). The simulation is run in 2D axisymmetric geometry. 
The simulations allowed for separate ion, electron, and radiation temperatures. The 6-ns temporally-tiled configuration was found by optimizing for sustained high midplane velocities and low collisionality. The predicted midplane single-flow velocity is shown in Fig.~\ref{fig:epwdata}e,
with comparisons to single-flow and two-flow Thomson scattering data. The qualitative agreement between the FLASH simulation (which was not 
% tuned for 
corrected to match
laser absorption, and was instead multiplied by 0.92 for qualitative comparison) and the measured velocity gives confidence that FLASH is doing a good job of predicting the single-flow plasma conditions. The FLASH simulation gives an estimate of the (not otherwise directly measured) arrival time of the flows at the experiment midplane, \qty{2.6}{\nano\second} (this estimate is made using the unscaled FLASH simulation).

\section{Proton radiography measurements}
\label{sec:prad}

Proton radiography provides a path-integrated measurement of electromagnetic fields in the interaction region and is able to record a 2D projection of the morphology of these field structures. 14.7 MeV protons from D$^3$He fusion in an implosion backlighter traverse the plasma and their spatial distribution is recorded on CR-39. After correcting for large-scale fluence variations, small-scale modulations in proton flux are interpreted as resulting primarily from magnetic deflections associated with transverse filament fields. 

\begin{figure*}
\includegraphics[width=\textwidth]{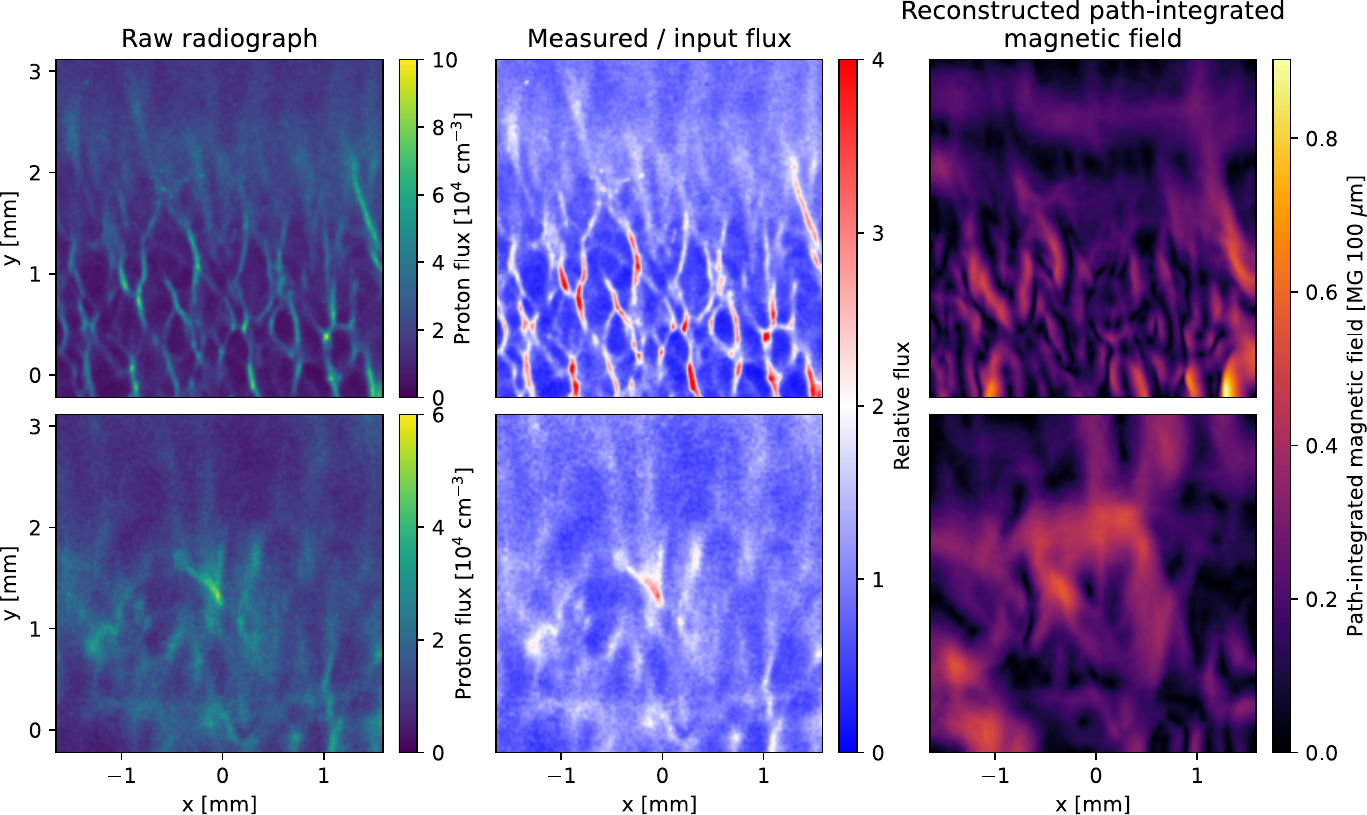}
\caption{\label{fig:pradrecon} Proton radiographs (first column) at 9 ns (top row) and 14 ns delay (bottom row). Radiographs  normalized to the background flux computed by low-pass filtering the image at a long length scale, $\sim$\qty{1}{\milli\meter} (second column). Magnetic field reconstructions of the proton radiographs, assuming all deflections are from magnetic fields (third column).}
\end{figure*}

Example radiographs are shown in Fig.~\ref{fig:pradrecon}. 
The preliminary radiographs collected and shown here were from a laser drive configuration using \qty{1}{\nano\second} pulses of the same total energy, \qty{5.4}{\kilo\joule} per foil as the Thomson-characterized systems.  
The D$^3$He source is located \qty{10}{mm} away from  TCC opposite the proton radiography detector stack, which is \qty{26}{cm} away from TCC. This corresponds to a magnification of 27, and a field-of-view at the subject plasma of \qty{3.7}{\milli\meter}.
The filamentary morphology seen here has been well-established by prior work \cite{ross_characterizing_2012,park_collisionless_2015,huntington_observation_2015} in similar conditions as being generated by the ion-Weibel instability. 
The filaments are expected to increase in size (and correspondingly decrease in magnetic field) as they coalesce with time (a result of flux conservation \cite{zhou_magnetic_2019}). The increase in structure size is qualitatively visible in the evolution from 9 ns to 14 ns (Fig.~\ref{fig:pradrecon}(a)-(c)).
Magnetic field reconstructions from these radiographs, calculated by solving a Monge-Ampere equation for the deflection field of the protons\cite{sulman_efficient_2011,bott_proton_2017}, show that the peak path-integrated fields observed at \qty{9}{ns} are of order \qty{50}{\mega\gauss\micro\meter}.  Assuming our proton distribution measurement is the result of a stochastic sampling of kicks from individual filaments, the reconstructed field should be distributed like a binomial distribution, so that:
\begin{equation*}
    BL_{max,observed} = \sqrt{N_{filaments}} B_{filament} L_{filament}
\end{equation*}
which can be rearranged for $B_{filament}$. The filament structure wavelength scale, in the 9 ns measurement, is approximately \qty{500}{\micro\meter}, a scale length confirmed by taking a Fourier transform of the proton image and selecting the peak wavelength, as has been previously demonstrated to be representative of typical filament length scale\cite{levesque_characterizing_2019}. Assuming an interaction region extent of \qty{3}{mm} (transverse to the flow, as seen in the radiograph), probe protons traverse $\sim$12 filaments as they pass the plasma (a "filament" size is $\frac{1}{2}$ of the structure wavelength). The result is that the typical filament magnetic field strength is $B_{filament}\sim$\qty{30}{\kilo\gauss} at \qty{9}{\nano\second}.  
The time evolution of the magnetic field and filament scales will be further explored in a future paper.

\section{Discussion}
\label{sec:discussion}

\begin{figure}
\includegraphics[width=\figcolwidth]{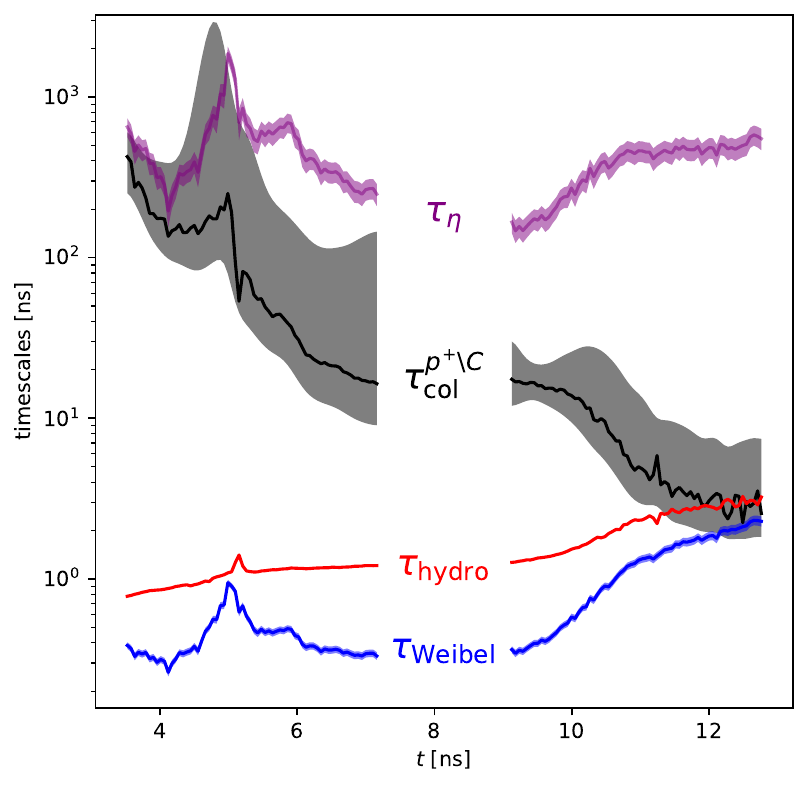}
\caption{\label{fig:timescales}
Relevant timescales, calculated using plasma conditions from the fits to Thomson scattering spectra. All timescales are shown with a shaded region indicating the typical standard error in the calculated quantities.
The resistive diffusion timescale $\tau_{\eta}$ (purple) is calculated from Drake\cite{drake_high-energy-density_2019} for lengthscales which match the fastest growing mode, which is a conservative estimate.
The ion inter-flow collision timescale $\tau^{p^{+}\backslash C}_{\mathrm{col}}$ (black) of protons-on-carbon is shown as it is the fastest collision timescale; the bottom (top) of the shaded region correspond to the different collisionality expeirenced by protons from lower- (higher-) density flows, while the solid line is calculated assuming a fixed 1:1 ratio of densities between flow components (appropriate for early-time pre-filament formation or late-time inter-filament regions).
The typical hydrodynamic timescale for the two flows to interpenetrate a distance of \qty{2}{\milli\meter}, $\tau_{\mathrm{hydro}}$, is also plotted (red). 
Finally, the linear growth rate of the fastest growing mode of the ion Weibel instability $\tau_{\mathrm{Weibel}}$ (blue) is plotted,calculated from Davidson \emph{et~al.}\cite{davidson_nonlinear_1972}. The fastest growing mode varies based on plasma conditions but spans the range 60-\qty{160}{\micro\meter}.
}
\end{figure}

\begin{figure}
\includegraphics[width=\figcolwidth]{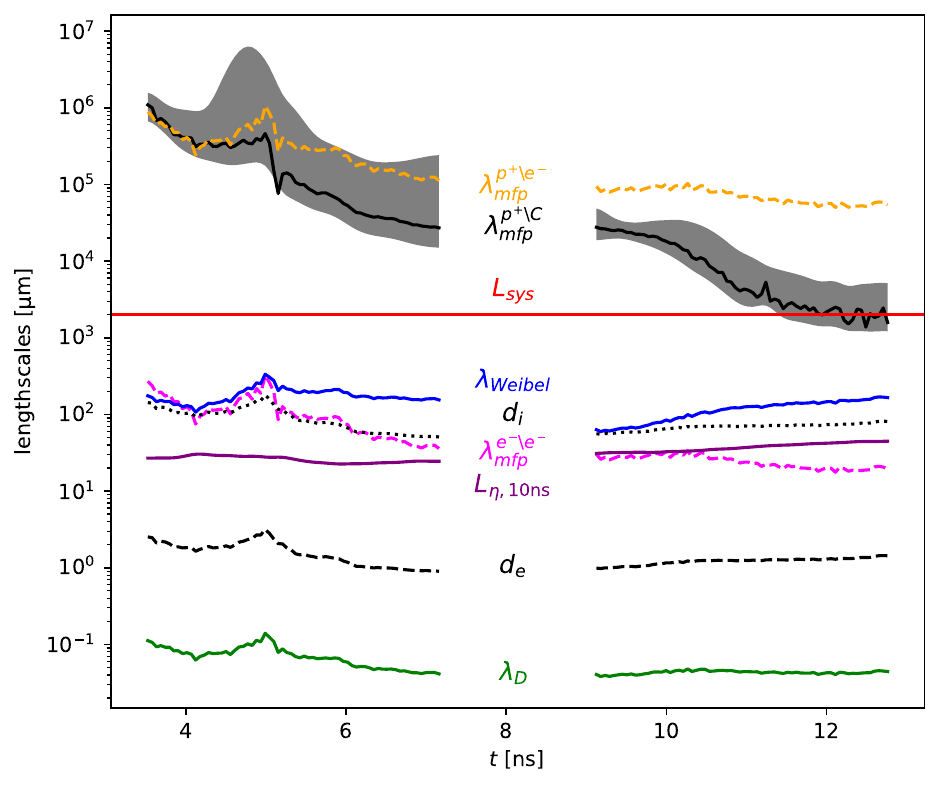}
\caption{\label{fig:lengthscales}
Heirarchy of relevant lengthscales, calculated using plasma conditions from the fits to Thomson scattering spectra (similar to Fig.~\ref{fig:timescales}). 
The resistive diffusion lengthscale, $L_{\eta,\mathrm{10 ns}}$, is the characteristic length over which fields can diffuse in 10 ns.
The ion inter-flow collision mean free paths of protons-on-carbon ($\lambda^{p^{+}\backslash C}_{mfp}$), protons-on-electrons ($\lambda^{p^{+}\backslash e^{-}}_{mfp}$), and electrons-on-electrons ($\lambda^{e^{-}\backslash e^{-}}_{mfp}$) are shown.
The typical system lengthscale of \qty{2}{\milli\meter} is shown (red).
The ion and electron inertial lengths are shown ($d_i$, $d_e$).
The wavelength of the fastest growing mode of the ion Weibel instability is shown $\lambda_{Weibel}$,calculated from Davidson \emph{et~al.}\cite{davidson_nonlinear_1972}
}
\end{figure}

The time-resolved Thomson scattering data shows that the interpenetrating flows are collisionless (i.e. ions of the majority flow do not experience substantial Coulomb collisions with ions/electrons of the other flow) until at least \qty{11.3}{\nano\second}, later than any previous OMEGA experiment. This can be seen by inspecting the collision time and lengthscales calculated from Thomson-measured plasma conditions. Relevant timescales are shown in Fig.~\ref{fig:timescales}: magnetic diffusion time $\tau_{\eta}$, proton-on-carbon Coulomb collision time $\tau^{p^{+}\backslash C}_{\mathrm{col}}$, characteristic volume flow crossing time, and the local linear ion Weibel growth time $\tau_{\mathrm{Weibel}}$. A comparison of relevant lengthscales (including typical inertial lengths, typical Weibel wavelength,system size, and mean free paths) is shown in Fig.~\ref{fig:lengthscales}. Both of these plots make clear that the collisional time/lengthscales are separated from the system scale until \qty{11.3}{\nano\second}. When the two-flow system begins decelerating faster than the one-flow system, at $\sim$\qty{9.5}{\nano\second}, the fastest collision timescale experienced by the flow ions is for protons-on-carbon, and is over \qty{10}{\nano\second} (and the mean free path is several \unit{\centi\meter}, much larger than the experiment size). 
The observed deceleration of the flows prior to this time must therefore be through the collisionless mechanisms; the generation of magnetic fields through the ion Weibel instability allows for the flows to interact.
Though the Weibel instability, the ion flows organize themselves into channels each with a dominant flow direction as seen in previous work\cite{swadling_measurement_2020,bruulsema_local_2020} and evident here from the asymmetry in flow density. 

As has been argued for similar past experiments, the ion Weibel instability observed here is well beyond the threshold of saturation. This can be seen by inspection of the saturation criterion from Davidson\cite{davidson_nonlinear_1972}, which indicates that the instability saturates when the ion bounce frequency $\omega_{i,bounce}$ grows to be of the same order as the ion Weibel linear growth rate $\gamma$: 
\begin{align*}
    \gamma &\sim k v \sqrt{\omega_{pi}^2/\omega_{pe}^2} \\
    \omega_{i,bounce} &\sim \sqrt{k v \frac{ZeB}{m_i c}}
\end{align*} \\
The above $\gamma$ expression applies when $k^2 \ll k_0^2$, where $k_0$ is the largest growing wavenumber (i.e. cutoff wavenumber). The typical bounce frequency for $k = k_0/3$ (satisfying $k^2 \ll k_0^2$ and close to the peak growing wavenumber) given our Thomson-measured plasma conditions and proton radiography estimated magnetic field at \qty{9}{\nano\second} is $\sim$\qty{200}{\pico\second}, which is less than the linear ion Weibel growth rate of $\sim$\qty{400}{\pico\second} indicating that the instability has already reached saturation. The linear growth saturates within a few nanoseconds of flow interpenetration, consistent with a few linear growth times and the findings of previous experiments\cite{manuel_experimental_2022,yuan_electron_2024}.

Resistive diffusion times of magnetic fields are also estimated directly from the measured two-flow Thomson parameters. From Drake\cite{drake_high-energy-density_2019}, 
\begin{equation*}
    \tau_{diff} = 12.8 \frac{L^2 T_e^{3/2}}{\bar{Z}\log \Lambda} \text{[ns]} 
\end{equation*}
with $L$ the typical magnetic field structure size in mm and $T_e$ the electron temperature in \unit{\electronvolt}. We expect the ion Weibel  filaments to grow at the ion inertial scale, $\sim$\qty{100}{\micro\meter}. The smallest observed filaments in these experiments (i.e., at/after \qty{5}{\nano\second}) are $\sim$\qty{200}{\micro\meter}. Using this as the lengthscale, we calculate the fastest diffusion times (for the smallest filaments) to range from 150-\qty{700}{\nano\second} as shown in Fig.~\ref{fig:timescales} which is significantly longer than the experimental evolution times. Thus we conclude that the resistive diffusion of magnetic fields does not play a major role in their evolution in this platform, which will be important in interpreting the morphology changes seen in the magnetic field reconstructions. 

\section{Conclusion and Future Work}
\label{sec:conclusion} 

We have developed an experimental platform which supports the ion Weibel instability, building on previous work at OMEGA and NIF. 
This platform sustains low-collisionality, low-resistivity, interpenetrating flows long enough to observe late-time 
%(multiple-generation) 
ion Weibel filament evolution, creating a unique experimental opportunity to discriminate between competing long-time filament-merger models. 
Our Thomson scattering measurements of the plasma conditions, and proton radiography measurements of typical field strength and morphology, support the following conclusions. 
The interpenetrating flows are collisionless until at least \qty{11.3}{\nano\second}, later than any previous OMEGA experiment. 
The observed deceleration of the flows must be through collisionless mechanisms; the generation of magnetic fields through the ion Weibel instability allows for the flows to interact.
Through the Weibel instability, the ion flows organize themselves into channels each with a dominant flow direction.
Path integrated magnetic fields are observed with magnitude \qty{50}{\mega\gauss\micro\meter} in the filaments which have wavelength \qty{500}{\micro\meter}, implying $\sim$\qty{30}{\kilo\gauss} filament fields.   
Resistive timescales are sufficiently long so as to not be important to the observed evolution of fields.

Further future interrogation of this data will explore the evolution of the spatial structure of the fields, and seek to understand the key physics at play as the saturated filaments interact with each other electromagnetically. Key questions include the possible role of magnetic reconnection in setting the limiting timescale of filament merger, and other instabilities in interrupting the otherwise flow-aligned filament morphology. Details about the partition of energy between ion and electron species should be possible to answer through careful analysis of the ion and electron temperatures in the Thomson scattering data.

\begin{acknowledgments}

The authors thank E.~Doeg and R.~Frankel (MIT) for their contributions in the etch/scan processing of the CR-39 data in this work. 
This work was performed under the auspices of the U.S. Department of Energy by Lawrence Livermore National Laboratory under Contract DE-AC52-07NA27344. 
Experimental time was supported in part by the National Nuclear Administration Program through the National Laser Users' Facility. 
M.Z. acknowledges support from the NSF Grant PHY.~2512037.
A.B. acknowledges support by the Spanish Ministerio de Ciencia, Innovación y Universidades (Grant No. PID2024-157933OA-I0).
The software used in this work was developed in part by the DOE NNSA- and DOE Office of Science-supported Flash Center for Computational Science at the University of Chicago and the University of Rochester.
LLNL-JRNL-2018338.
% This article has been submitted to by Physics of Plasmas. After it is published, it will be found at \href{https://publishing.aip.org/resources/librarians/products/journals/}{Link}.

\end{acknowledgments}

% \section*{Data Availability Statement}

% \appendix

% \section{Appendixes}

% \nocite{*}
% \bibliography{counterstreaming.bib}% Produces the bibliography via BibTeX.
\bibliography{IonWeibelfilaments.bib}% Produces the bibliography via BibTeX.

@article{sulman_efficient_2011,
	title = {An efficient approach for the numerical solution of the {Monge}–{Ampère} equation},
	volume = {61},
	copyright = {https://www.elsevier.com/tdm/userlicense/1.0/},
	issn = {01689274},
	url = {https://linkinghub.elsevier.com/retrieve/pii/S0168927410001819},
	doi = {10.1016/j.apnum.2010.10.006},
	language = {en},
	number = {3},
	urldate = {2025-10-31},
	journal = {Applied Numerical Mathematics},
	author = {Sulman, Mohamed M. and Williams, J.F. and Russell, Robert D.},
	month = mar,
	year = {2011},
	pages = {298--307},
}

@article{davidson_nonlinear_1972,
	title = {Nonlinear {Development} of {Electromagnetic} {Instabilities} in {Anisotropic} {Plasmas}},
	volume = {15},
	issn = {0031-9171},
	url = {https://pubs.aip.org/pfl/article/15/2/317/448497/Nonlinear-Development-of-Electromagnetic},
	doi = {10.1063/1.1693910},
	language = {en},
	number = {2},
	urldate = {2025-10-31},
	journal = {The Physics of Fluids},
	author = {Davidson, Ronald C. and Hammer, David A. and Haber, Irving and Wagner, Carl E.},
	month = feb,
	year = {1972},
	pages = {317--333},
}

@article{fiuza_electron_2020,
	title = {Electron acceleration in laboratory-produced turbulent collisionless shocks},
	volume = {16},
	issn = {1745-2473, 1745-2481},
	url = {https://www.nature.com/articles/s41567-020-0919-4},
	doi = {10.1038/s41567-020-0919-4},
	language = {en},
	number = {9},
	urldate = {2025-10-31},
	journal = {Nature Physics},
	author = {Fiuza, F. and Swadling, G. F. and Grassi, A. and Rinderknecht, H. G. and Higginson, D. P. and Ryutov, D. D. and Bruulsema, C. and Drake, R. P. and Funk, S. and Glenzer, S. and Gregori, G. and Li, C. K. and Pollock, B. B. and Remington, B. A. and Ross, J. S. and Rozmus, W. and Sakawa, Y. and Spitkovsky, A. and Wilks, S. and Park, H.-S.},
	month = sep,
	year = {2020},
	pages = {916--920},
}

@article{kasim_quantitative_2017,
	title = {Quantitative shadowgraphy and proton radiography for large intensity modulations},
	volume = {95},
	issn = {2470-0045, 2470-0053},
	url = {http://arxiv.org/abs/1607.04179},
	doi = {10.1103/PhysRevE.95.023306},
	language = {en},
	number = {2},
	urldate = {2025-10-31},
	journal = {Physical Review E},
	author = {Kasim, Muhammad Firmansyah and Ceurvorst, Luke and Ratan, Naren and Sadler, James and Chen, Nicholas and Savert, Alexander and Trines, Raoul and Bingham, Robert and Burrows, Philip N. and Kaluza, Malte C. and Norreys, Peter},
	month = feb,
	year = {2017},
	note = {arXiv:1607.04179 [physics]},
	pages = {023306},
}

@article{bruulsema_local_2020,
	title = {On the local measurement of electric currents and magnetic fields using {Thomson} scattering in {Weibel}-unstable plasmas},
	volume = {27},
	issn = {1070-664X, 1089-7674},
	url = {https://pubs.aip.org/pop/article/27/5/052104/290555/On-the-local-measurement-of-electric-currents-and},
	doi = {10.1063/1.5140674},
	language = {en},
	number = {5},
	urldate = {2025-10-31},
	journal = {Physics of Plasmas},
	author = {Bruulsema, C. and Rozmus, W. and Swadling, G. F. and Glenzer, S. and Park, H. S. and Ross, J. S. and Fiuza, F.},
	month = may,
	year = {2020},
	pages = {052104},
}

@article{ross_characterizing_2012,
	title = {Characterizing counter-streaming interpenetrating plasmas relevant to astrophysical collisionless shocks},
	volume = {19},
	issn = {1070-664X, 1089-7674},
	url = {https://pubs.aip.org/pop/article/19/5/056501/596990/Characterizing-counter-streaming-interpenetrating},
	doi = {10.1063/1.3694124},
	language = {en},
	number = {5},
	urldate = {2025-10-31},
	journal = {Physics of Plasmas},
	author = {Ross, J. S. and Glenzer, S. H. and Amendt, P. and Berger, R. and Divol, L. and Kugland, N. L. and Landen, O. L. and Plechaty, C. and Remington, B. and Ryutov, D. and Rozmus, W. and Froula, D. H. and Fiksel, G. and Sorce, C. and Kuramitsu, Y. and Morita, T. and Sakawa, Y. and Takabe, H. and Drake, R. P. and Grosskopf, M. and Kuranz, C. and Gregori, G. and Meinecke, J. and Murphy, C. D. and Koenig, M. and Pelka, A. and Ravasio, A. and Vinci, T. and Liang, E. and Presura, R. and Spitkovsky, A. and Miniati, F. and Park, H.-S.},
	month = may,
	year = {2012},
	pages = {056501},
}

@article{swadling_measurement_2020,
	title = {Measurement of {Kinetic}-{Scale} {Current} {Filamentation} {Dynamics} and {Associated} {Magnetic} {Fields} in {Interpenetrating} {Plasmas}},
	volume = {124},
	issn = {0031-9007, 1079-7114},
	url = {https://link.aps.org/doi/10.1103/PhysRevLett.124.215001},
	doi = {10.1103/PhysRevLett.124.215001},
	language = {en},
	number = {21},
	urldate = {2025-10-31},
	journal = {Physical Review Letters},
	author = {Swadling, G. F. and Bruulsema, C. and Fiuza, F. and Higginson, D. P. and Huntington, C. M. and Park, H-S. and Pollock, B. B. and Rozmus, W. and Rinderknecht, H. G. and Katz, J. and Birkel, A. and Ross, J. S.},
	month = may,
	year = {2020},
	pages = {215001},
}

@article{kulsrud_origin_2008,
	title = {On the origin of cosmic magnetic fields},
	volume = {71},
	issn = {0034-4885, 1361-6633},
	url = {https://iopscience.iop.org/article/10.1088/0034-4885/71/4/046901},
	doi = {10.1088/0034-4885/71/4/046901},
	language = {en},
	number = {4},
	urldate = {2025-11-03},
	journal = {Reports on Progress in Physics},
	author = {Kulsrud, Russell M and Zweibel, Ellen G},
	month = apr,
	year = {2008},
	pages = {046901},
}

@article{weibel_spontaneously_1959,
	title = {Spontaneously {Growing} {Transverse} {Waves} in a {Plasma} {Due} to an {Anisotropic} {Velocity} {Distribution}},
	volume = {2},
	copyright = {http://link.aps.org/licenses/aps-default-license},
	issn = {0031-9007},
	url = {https://link.aps.org/doi/10.1103/PhysRevLett.2.83},
	doi = {10.1103/PhysRevLett.2.83},
	language = {en},
	number = {3},
	urldate = {2025-11-03},
	journal = {Physical Review Letters},
	author = {Weibel, Erich S.},
	month = feb,
	year = {1959},
	pages = {83--84},
}

@article{manuel_experimental_2022,
	title = {Experimental evidence of early-time saturation of the ion-{Weibel} instability in counterstreaming plasmas of {CH}, {Al}, and {Cu}},
	volume = {106},
	issn = {2470-0045, 2470-0053},
	url = {https://link.aps.org/doi/10.1103/PhysRevE.106.055205},
	doi = {10.1103/PhysRevE.106.055205},
	language = {en},
	number = {5},
	urldate = {2025-11-10},
	journal = {Physical Review E},
	author = {Manuel, M. J.-E. and Adams, M. B. P. and Ghosh, S. and Beg, F. N. and Bolaños, S. and Huntington, C. M. and Jonnalagadda, R. and Kawahito, D. and Pollock, B. B. and Remington, B. A. and Ross, J. S. and Ryutov, D. D. and Sio, H. and Swadling, G. F. and Tzeferacos, P. and Park, H.-S.},
	month = nov,
	year = {2022},
	pages = {055205},
}

@article{huntington_observation_2015,
	title = {Observation of magnetic field generation via the {Weibel} instability in interpenetrating plasma flows},
	volume = {11},
	issn = {1745-2473, 1745-2481},
	url = {https://www.nature.com/articles/nphys3178},
	doi = {10.1038/nphys3178},
	language = {en},
	number = {2},
	urldate = {2025-11-10},
	journal = {Nature Physics},
	author = {Huntington, C. M. and Fiuza, F. and Ross, J. S. and Zylstra, A. B. and Drake, R. P. and Froula, D. H. and Gregori, G. and Kugland, N. L. and Kuranz, C. C. and Levy, M. C. and Li, C. K. and Meinecke, J. and Morita, T. and Petrasso, R. and Plechaty, C. and Remington, B. A. and Ryutov, D. D. and Sakawa, Y. and Spitkovsky, A. and Takabe, H. and Park, H.-S.},
	month = feb,
	year = {2015},
	pages = {173--176},
}

@article{huntington_magnetic_2017,
	title = {Magnetic field production via the {Weibel} instability in interpenetrating plasma flows},
	volume = {24},
	issn = {1070-664X, 1089-7674},
	url = {https://pubs.aip.org/pop/article/24/4/041410/110100/Magnetic-field-production-via-the-Weibel},
	doi = {10.1063/1.4982044},
	language = {en},
	number = {4},
	urldate = {2025-11-10},
	journal = {Physics of Plasmas},
	author = {Huntington, C. M. and Manuel, M. J.-E. and Ross, J. S. and Wilks, S. C. and Fiuza, F. and Rinderknecht, H. G. and Park, H.-S. and Gregori, G. and Higginson, D. P. and Park, J. and Pollock, B. B. and Remington, B. A. and Ryutov, D. D. and Ruyer, C. and Sakawa, Y. and Sio, H. and Spitkovsky, A. and Swadling, G. F. and Takabe, H. and Zylstra, A. B.},
	month = apr,
	year = {2017},
	pages = {041410},
}

@article{park_collisionless_2015,
	title = {Collisionless shock experiments with lasers and observation of {Weibel} instabilities},
	volume = {22},
	issn = {1070-664X, 1089-7674},
	url = {https://pubs.aip.org/pop/article/22/5/056311/110799/Collisionless-shock-experiments-with-lasers-and},
	doi = {10.1063/1.4920959},
	language = {en},
	number = {5},
	urldate = {2025-11-10},
	journal = {Physics of Plasmas},
	author = {Park, H.-S. and Huntington, C. M. and Fiuza, F. and Drake, R. P. and Froula, D. H. and Gregori, G. and Koenig, M. and Kugland, N. L. and Kuranz, C. C. and Lamb, D. Q. and Levy, M. C. and Li, C. K. and Meinecke, J. and Morita, T. and Petrasso, R. D. and Pollock, B. B. and Remington, B. A. and Rinderknecht, H. G. and Rosenberg, M. and Ross, J. S. and Ryutov, D. D. and Sakawa, Y. and Spitkovsky, A. and Takabe, H. and Turnbull, D. P. and Tzeferacos, P. and Weber, S. V. and Zylstra, A. B.},
	month = may,
	year = {2015},
	pages = {056311},
}

@article{zhou_magnetic_2019,
	title = {Magnetic island merger as a mechanism for inverse magnetic energy transfer},
	volume = {1},
	issn = {2643-1564},
	url = {https://link.aps.org/doi/10.1103/PhysRevResearch.1.012004},
	doi = {10.1103/PhysRevResearch.1.012004},
	language = {en},
	number = {1},
	urldate = {2025-11-10},
	journal = {Physical Review Research},
	author = {Zhou, Muni and Bhat, Pallavi and Loureiro, Nuno F. and Uzdensky, Dmitri A.},
	month = aug,
	year = {2019},
	pages = {012004},
}

@article{ruyer_nonlinear_2015,
	title = {Nonlinear dynamics of the ion {Weibel}-filamentation instability: {An} analytical model for the evolution of the plasma and spectral properties},
	volume = {22},
	issn = {1070-664X, 1089-7674},
	shorttitle = {Nonlinear dynamics of the ion {Weibel}-filamentation instability},
	url = {https://pubs.aip.org/pop/article/22/3/032102/929092/Nonlinear-dynamics-of-the-ion-Weibel-filamentation},
	doi = {10.1063/1.4913651},
	language = {en},
	number = {3},
	urldate = {2025-11-10},
	journal = {Physics of Plasmas},
	author = {Ruyer, C. and Gremillet, L. and Debayle, A. and Bonnaud, G.},
	month = mar,
	year = {2015},
	pages = {032102},
}

@article{schoeffler_magnetic-field_2014,
	title = {Magnetic-{Field} {Generation} and {Amplification} in an {Expanding} {Plasma}},
	volume = {112},
	copyright = {http://link.aps.org/licenses/aps-default-license},
	issn = {0031-9007, 1079-7114},
	url = {https://link.aps.org/doi/10.1103/PhysRevLett.112.175001},
	doi = {10.1103/PhysRevLett.112.175001},
	language = {en},
	number = {17},
	urldate = {2025-11-10},
	journal = {Physical Review Letters},
	author = {Schoeffler, K. M. and Loureiro, N. F. and Fonseca, R. A. and Silva, L. O.},
	month = apr,
	year = {2014},
	pages = {175001},
}

@article{schoeffler_generation_2016,
	title = {The generation of magnetic fields by the {Biermann} battery and the interplay with the {Weibel} instability},
	volume = {23},
	issn = {1070-664X, 1089-7674},
	url = {https://pubs.aip.org/pop/article/23/5/056304/966354/The-generation-of-magnetic-fields-by-the-Biermann},
	doi = {10.1063/1.4946017},
	language = {en},
	number = {5},
	urldate = {2025-11-10},
	journal = {Physics of Plasmas},
	author = {Schoeffler, K. M. and Loureiro, N. F. and Fonseca, R. A. and Silva, L. O.},
	month = may,
	year = {2016},
	pages = {056304},
}

@article{li_monoenergetic_2006,
	title = {Monoenergetic proton backlighter for measuring {E} and {B} fields and for radiographing implosions and high-energy density plasmas (invited)},
	volume = {77},
	issn = {0034-6748, 1089-7623},
	url = {https://pubs.aip.org/rsi/article/77/10/10E725/1017998/Monoenergetic-proton-backlighter-for-measuring-E},
	doi = {10.1063/1.2228252},
	language = {en},
	number = {10},
	urldate = {2026-03-04},
	journal = {Review of Scientific Instruments},
	author = {Li, C. K. and Séguin, F. H. and Frenje, J. A. and Rygg, J. R. and Petrasso, R. D. and Town, R. P. J. and Amendt, P. A. and Hatchett, S. P. and Landen, O. L. and Mackinnon, A. J. and Patel, P. K. and Smalyuk, V. A. and Knauer, J. P. and Sangster, T. C. and Stoeckl, C.},
	month = oct,
	year = {2006},
	pages = {10E725},
}

@article{sulman_efficient_2011-1,
	title = {An efficient approach for the numerical solution of the {Monge}–{Ampère} equation},
	volume = {61},
	copyright = {https://www.elsevier.com/tdm/userlicense/1.0/},
	issn = {01689274},
	url = {https://linkinghub.elsevier.com/retrieve/pii/S0168927410001819},
	doi = {10.1016/j.apnum.2010.10.006},
	language = {en},
	number = {3},
	urldate = {2026-03-11},
	journal = {Applied Numerical Mathematics},
	author = {Sulman, Mohamed M. and Williams, J.F. and Russell, Robert D.},
	month = mar,
	year = {2011},
	pages = {298--307},
}

@article{bott_proton_2017,
	title = {Proton imaging of stochastic magnetic fields},
	volume = {83},
	copyright = {https://www.cambridge.org/core/terms},
	issn = {0022-3778, 1469-7807},
	url = {https://www.cambridge.org/core/product/identifier/S0022377817000939/type/journal_article},
	doi = {10.1017/S0022377817000939},
	language = {en},
	number = {6},
	urldate = {2026-03-14},
	journal = {Journal of Plasma Physics},
	author = {Bott, A. F. A. and Graziani, C. and Tzeferacos, P. and White, T. G. and Lamb, D. Q. and Gregori, G. and Schekochihin, A. A.},
	month = dec,
	year = {2017},
	pages = {905830614},
}

@article{medvedev_long-time_2005,
	title = {Long-{Time} {Evolution} of {Magnetic} {Fields} in {Relativistic} {Gamma}-{Ray} {Burst} {Shocks}},
	volume = {618},
	issn = {0004-637X, 1538-4357},
	url = {https://iopscience.iop.org/article/10.1086/427921},
	doi = {10.1086/427921},
	language = {en},
	number = {2},
	urldate = {2026-04-16},
	journal = {The Astrophysical Journal},
	author = {Medvedev, Mikhail V. and Fiore, Massimiliano and Fonseca, Ricardo A. and Silva, Luis O. and Mori, Warren B.},
	month = jan,
	year = {2005},
	pages = {L75--L78},
}

@article{takabe_theory_2023,
	title = {Theory of magnetic turbulence and shock formation induced by a collisionless plasma instability},
	volume = {30},
	issn = {1070-664X, 1089-7674},
	url = {https://pubs.aip.org/pop/article/30/3/030901/2881653/Theory-of-magnetic-turbulence-and-shock-formation},
	doi = {10.1063/5.0130264},
	language = {en},
	number = {3},
	urldate = {2026-04-16},
	journal = {Physics of Plasmas},
	author = {Takabe, Hideaki},
	month = mar,
	year = {2023},
	pages = {030901},
}

@article{ruyer_disruption_2018,
	title = {Disruption of {Current} {Filaments} and {Isotropization} of the {Magnetic} {Field} in {Counterstreaming} {Plasmas}},
	volume = {120},
	issn = {0031-9007, 1079-7114},
	url = {https://link.aps.org/doi/10.1103/PhysRevLett.120.245002},
	doi = {10.1103/PhysRevLett.120.245002},
	language = {en},
	number = {24},
	urldate = {2026-04-16},
	journal = {Physical Review Letters},
	author = {Ruyer, C. and Fiuza, F.},
	month = jun,
	year = {2018},
	pages = {245002},
}

@article{marret_energy_2026-1,
	title = {Energy {Partition} in {Collisionless} {Counterstreaming} {Plasmas}},
	volume = {997},
	issn = {2041-8205, 2041-8213},
	url = {https://iopscience.iop.org/article/10.3847/2041-8213/ae3060},
	doi = {10.3847/2041-8213/ae3060},
	language = {en},
	number = {1},
	urldate = {2026-04-16},
	journal = {The Astrophysical Journal Letters},
	author = {Marret, Alexis and Fiuza, Frederico},
	month = jan,
	year = {2026},
	pages = {L23},
}

@article{seguin_spectrometry_2003,
	title = {Spectrometry of charged particles from inertial-confinement-fusion plasmas},
	volume = {74},
	issn = {0034-6748, 1089-7623},
	url = {https://pubs.aip.org/rsi/article/74/2/975/348627/Spectrometry-of-charged-particles-from-inertial},
	doi = {10.1063/1.1518141},
	language = {en},
	number = {2},
	urldate = {2026-04-16},
	journal = {Review of Scientific Instruments},
	author = {Séguin, F. H. and Frenje, J. A. and Li, C. K. and Hicks, D. G. and Kurebayashi, S. and Rygg, J. R. and Schwartz, B.-E. and Petrasso, R. D. and Roberts, S. and Soures, J. M. and Meyerhofer, D. D. and Sangster, T. C. and Knauer, J. P. and Sorce, C. and Glebov, V. Yu. and Stoeckl, C. and Phillips, T. W. and Leeper, R. J. and Fletcher, K. and Padalino, S.},
	month = feb,
	year = {2003},
	pages = {975--995},
}

@article{ross_transition_2017,
	title = {Transition from {Collisional} to {Collisionless} {Regimes} in {Interpenetrating} {Plasma} {Flows} on the {National} {Ignition} {Facility}},
	volume = {118},
	copyright = {http://link.aps.org/licenses/aps-default-license},
	issn = {0031-9007, 1079-7114},
	url = {http://link.aps.org/doi/10.1103/PhysRevLett.118.185003},
	doi = {10.1103/PhysRevLett.118.185003},
	language = {en},
	number = {18},
	urldate = {2026-04-16},
	journal = {Physical Review Letters},
	author = {Ross, J. S. and Higginson, D. P. and Ryutov, D. and Fiuza, F. and Hatarik, R. and Huntington, C. M. and Kalantar, D. H. and Link, A. and Pollock, B. B. and Remington, B. A. and Rinderknecht, H. G. and Swadling, G. F. and Turnbull, D. P. and Weber, S. and Wilks, S. and Froula, D. H. and Rosenberg, M. J. and Morita, T. and Sakawa, Y. and Takabe, H. and Drake, R. P. and Kuranz, C. and Gregori, G. and Meinecke, J. and Levy, M. C. and Koenig, M. and Spitkovsky, A. and Petrasso, R. D. and Li, C. K. and Sio, H. and Lahmann, B. and Zylstra, A. B. and Park, H.-S.},
	month = may,
	year = {2017},
	pages = {185003},
}

@article{levesque_characterizing_2019,
	title = {Characterizing filamentary magnetic structures in counter-streaming plasmas by {Fourier} analysis of proton images},
	volume = {26},
	issn = {1070-664X, 1089-7674},
	url = {https://pubs.aip.org/pop/article/26/10/102303/264005/Characterizing-filamentary-magnetic-structures-in},
	doi = {10.1063/1.5100728},
	language = {en},
	number = {10},
	urldate = {2026-04-16},
	journal = {Physics of Plasmas},
	author = {Levesque, Joseph and Kuranz, Carolyn and Handy, Timothy and Manuel, Mario and Fiuza, Frederico},
	month = oct,
	year = {2019},
	pages = {102303},
}

@book{drake_high-energy-density_2019,
	address = {Cham},
	edition = {Second edition, first softcover printing},
	series = {Graduate texts in physics},
	title = {High-energy-density physics: foundation of inertial fusion and experimental astrophysics},
	isbn = {978-3-319-67711-8 978-3-319-88473-8},
	shorttitle = {High-energy-density physics},
	publisher = {Springer},
	author = {Drake, R. Paul},
	year = {2019},
}

@article{fried_mechanism_1959,
	title = {Mechanism for {Instability} of {Transverse} {Plasma} {Waves}},
	volume = {2},
	issn = {0031-9171},
	url = {https://pubs.aip.org/pfl/article/2/3/337/942822/Mechanism-for-Instability-of-Transverse-Plasma},
	doi = {10.1063/1.1705933},
	language = {en},
	number = {3},
	urldate = {2026-05-08},
	journal = {The Physics of Fluids},
	author = {Fried, Burton D.},
	month = may,
	year = {1959},
	pages = {337--337},
}

@article{yuan_electron_2024,
	title = {Electron stochastic acceleration in laboratory-produced kinetic turbulent plasmas},
	volume = {15},
	issn = {2041-1723},
	url = {https://www.nature.com/articles/s41467-024-50085-7},
	doi = {10.1038/s41467-024-50085-7},
	language = {en},
	number = {1},
	urldate = {2026-05-08},
	journal = {Nature Communications},
	author = {Yuan, Dawei and Lei, Zhu and Wei, Huigang and Zhang, Zhe and Zhong, Jiayong and Li, Yifei and Ping, Yongli and Zhang, Yihang and Li, Yutong and Wang, Feilu and Liang, Guiyun and Qiao, Bin and Fu, Changbo and Liu, Huiya and Zhang, Panzheng and Zhu, Jianqiang and Zhao, Gang and Zhang, Jie},
	month = jul,
	year = {2024},
	pages = {5897},
}

@article{yuan_laboratory_2018,
	title = {Laboratory study of astrophysical collisionless shock at {SG}-{II} laser facility},
	volume = {6},
	copyright = {http://creativecommons.org/licenses/by/4.0/},
	issn = {2095-4719, 2052-3289},
	url = {https://www.cambridge.org/core/product/identifier/S2095471918000403/type/journal_article},
	doi = {10.1017/hpl.2018.40},
	language = {en},
	urldate = {2026-05-08},
	journal = {High Power Laser Science and Engineering},
	author = {Yuan, Dawei and Wei, Huigang and Liang, Guiyun and Wang, Feilu and Li, Yutong and Zhang, Zhe and Zhu, Baojun and Zhao, Jiarui and Jiang, Weiman and Han, Bo and Yuan, Xiaoxia and Zhong, Jiayong and Yuan, Xiaohui and Fu, Changbo and Zhang, Xiaopeng and Wang, Chen and Jia, Guo and Xiong, Jun and Fang, Zhiheng and Jiang, Shaoen and Du, Kai and Ding, Yongkun and Hua, Neng and Qiao, Zhanfeng and Zhou, Shenlei and Zhu, Baoqiang and Zhu, Jianqiang and Zhao, Gang and Zhang, Jie},
	year = {2018},
	pages = {e45},
}

@article{tzeferacos_flash_2015,
	title = {{FLASH} {MHD} simulations of experiments that study shock-generated magnetic fields},
	volume = {17},
	issn = {15741818},
	url = {https://linkinghub.elsevier.com/retrieve/pii/S1574181814000779},
	doi = {10.1016/j.hedp.2014.11.003},
	language = {en},
	urldate = {2026-05-08},
	journal = {High Energy Density Physics},
	author = {Tzeferacos, P. and Fatenejad, M. and Flocke, N. and Graziani, C. and Gregori, G. and Lamb, D.Q. and Lee, D. and Meinecke, J. and Scopatz, A. and Weide, K.},
	month = dec,
	year = {2015},
	pages = {24--31},
}

@article{fryxell_flash_2000,
	title = {{FLASH}: {An} {Adaptive} {Mesh} {Hydrodynamics} {Code} for {Modeling} {Astrophysical} {Thermonuclear} {Flashes}},
	volume = {131},
	url = {https://doi.org/10.1086/317361},
	doi = {10.1086/317361},
	number = {1},
	journal = {The Astrophysical Journal Supplement Series},
	author = {Fryxell, B. and Olson, K. and Ricker, P. and Timmes, F. X. and Zingale, M. and Lamb, D. Q. and MacNeice, P. and Rosner, R. and Truran, J. W. and Tufo, H.},
	month = nov,
	year = {2000},
	pages = {273},
}

@article{boehly_upgrade_1995,
	title = {The upgrade to the {OMEGA} laser system},
	volume = {66},
	issn = {0034-6748},
	url = {https://doi.org/10.1063/1.1146333},
	doi = {10.1063/1.1146333},
	number = {1},
	journal = {Review of Scientific Instruments},
	author = {Boehly, T. R. and Craxton, R. S. and Hinterman, T. H. and Kelly, J. H. and Kessler, T. J. and Kumpan, S. A. and Letzring, S. A. and McCrory, R. L. and Morse, S. F. B. and Seka, W. and Skupsky, S. and Soures, J. M. and Verdon, C. P.},
	month = jan,
	year = {1995},
	note = {\_eprint: https://pubs.aip.org/aip/rsi/article-pdf/66/1/508/19254304/508\_1\_online.pdf},
	pages = {508--510},
}

\end{document}